# Individual and Social Behaviour in Particle Swarm Optimizers

J. Sienz[1] and M. S. Innocente[1]
[1]ADOPT Research Group,
School of Engineering
Swansea University,
Swansea, UK

## Abstract

Three basic factors govern the individual behaviour of a particle: the *inertia from its previous displacement*; the *attraction to its own best experience*; and the *attraction to a given neighbour's best experience*. The importance awarded to each factor is controlled by three coefficients: the *inertia*; the *individuality*; and the *sociality weights*. The social behaviour is ruled by the structure of the social network, which defines the neighbours that are to inform of their experiences to a given particle. This paper presents a study of the influence of different settings of the coefficients as well as of the combined effect of different settings and different neighbourhood topologies on the speed and form of convergence.

**Keywords:** particle swarm optimization, velocity coefficients, neighbourhoods' topology, convergence.

## 1 Introduction

Particle Swarm Optimization (PSO) is a global optimizer in the sense that it is able to escape poor suboptimal solutions. This is possible thanks to a parallel search carried out by a population of cooperative individuals –called particles– which profit from sharing information acquired through experience. Thus, while Evolutionary Algorithms (EAs) rely on *competition* to improve the average fitness of a population (survival of the fittest), PSO relies on *cooperation*.

Individually, each particle is pulled by two attractors, while also carrying some inertia from its previous displacement. One of the attractors is its own best previous experience, and the other is the best previous experience of a given neighbour. Thus, these are the three basic ingredients ruling a particle's trajectory: the *inertia from its previous displacement*; the *attraction to its own best experience*; and the *attraction to a given neighbour's best experience*. The importance granted to each of these three ingredients is regulated by three coefficients: the *inertia*; the *individuality*; and the *sociality weights*. The individual behaviour of a particle is governed by the settings of these coefficients. Loosely speaking, a high inertia weight results in higher reluctance to changing the direction of its displacement; a high individuality weight results in higher confidence thus typically delaying convergence; and a high sociality weight leads to higher conformism thus typically accelerating convergence. There are, however, other issues for different combinations of settings that may modify this behaviour. For instance, increasing individuality over sociality may actually increase convergence speed, while some coefficients result in the particles diverging rather than clustering. In a similar manner as mutation does in EAs, random weights embedded in the particles' velocity update equation introduce creativity into the system so as to avoid getting trapped in some regular pattern.

The other important aspect with regards to the particles' behaviour is which neighbours are to inform of their best experiences to which particles. In other words, how to define the social attractor in the particles' velocity update equation, thus governing the particles' social behaviour. This leads to the development of infinite designs of social networks within the so-called *swarm* (population), which are typically referred to as *neighbourhood structures* or *neighbourhood topologies*. The latter must guarantee that information is spread throughout the whole swarm either at every time-step or at least eventually. Loosely speaking, the lower the number of interconnections the higher the degree of locality and hence the slower the convergence. Another means of affecting the social behaviour is by changing the number of attractors (e.g. having two social attractors −one local and one global− or by means of the so-called fully-informed PSO in [1] and [2]).

There is always the need of a trade-off between the explorative and the exploitative behaviour of the particles in the swarm. Explorative behaviour is more reluctant to getting trapped in suboptimal solutions whereas exploitative behaviour is better for a fine-grain search. This trade-off may be controlled by both the coefficients' settings and the neighbourhoods' topology. This paper presents a study of the former, and of the combination of some selected settings with three classical neighbourhood topologies, namely the *global*, the *wheel*, and the *ring* topologies (the latter with two neighbours only).

The remainder of this paper is organized as follows: a brief overview of the PSO method is provided in section 2; the study of the influence of the coefficients' settings is presented in section 3, where section 3.1 considers a single isolated particle and section 3.2 considers a small swarm of four interacting particles; the types of neighbourhoods and their interactions with the coefficients' settings to be considered in the experiments are discussed in section 4; and the experimental results are shown in section 5. Finally, some conclusions are offered in section 6.

## 2    Particle Swarm Optimization

Particle Swarm Optimization is a population-based and gradient-free optimization method introduced by social-psychologist James Kennedy and electric engineer Russell C. Eberhart in 1995 [3]. The method was inspired by earlier bird-flock simulations, and influenced by EAs. Therefore the method has roots on different fields, such as social psychology, Artificial Intelligence (AI), and mathematical optimization. Currently, its main applications are in solving optimization problems that are difficult to be handled by traditional methods. The algorithm is especially suitable for nonlinear problems with real-valued variables, although adaptations can be found in the literature to deal with discrete problems (e.g. [4, pp. 289-299]; [5]; [6]; and [7]). Since gradient information is not required, nondifferentiable and even discontinuous problems can be handled. In fact, given that the method imposes no restrictions to the functions involved, they do not even need to be explicit.

Since PSO is not deterministically implemented to optimize but to simulate some social behaviour, its optimization ability is an emergent property resulting from local interactions among the particles. This makes it difficult to understand its theoretical bases. Nonetheless, considerable theoretical work has been carried out on simplified versions of the algorithm (e.g. [8], [9], [4], [10], [11], [12], and [13]). For a comprehensive review of the PSO method, refer to [14] and [15].

## 2.1 Basic algorithm

While the ability to optimize emerges from decentralized local interactions among the particles in the swarm, the individual behaviour of each particle is governed by Equation (1):

$$\begin{cases} v_{ij}^{(t)} = w \cdot v_{ij}^{(t-1)} + iw \cdot U_{(0,1)} \cdot \left( pbest_{ij}^{(t-1)} - x_{ij}^{(t-1)} \right) + sw \cdot U_{(0,1)} \cdot \left( lbest_{ij}^{(t-1)} - x_{ij}^{(t-1)} \right) \\ x_{ij}^{(t)} = x_{ij}^{(t-1)} + v_{ij}^{(t)} \end{cases} \tag{1}$$

where:

| | | |
|---|---|---|
| $v_{ij}^{(t)}$ | : | $j^{\text{th}}$ component of the velocity of particle $i$ at time-step $t$. |
| $x_{ij}^{(t)}$ | : | $j^{\text{th}}$ coordinate of the position of particle $i$ at time-step $t$. |
| $U_{(0,1)}$ | : | Random number from a uniform distribution in the range [0,1] resampled anew every time it is referenced. |
| $w, iw, sw$ | : | Inertia, individuality, and sociality weights, respectively. |
| $pbest_{ij}^{(t)}$ | : | $j^{\text{th}}$ coordinate of the best position found by particle $i$ by time-step $t$. |
| $lbest_{ij}^{(t)}$ | : | $j^{\text{th}}$ coordinate of the best position found by any particle in the neighbourhood of particle $i$ by time-step $t$. |

Thus, the particle evaluates its current position in terms of a so-called *conflict function*, which is to be minimized (for minimization problems). Its performance is compared to its own best previous experience (**pbest**) and to the best previous experience within its neighbourhood (**lbest**). When moving to its next position, the particle is attracted by both experiences to some extent. There are three coefficients that govern the dynamics of the swarm: the *inertia* ($w$), the *individuality* ($iw$), and the *sociality* ($sw$) *weights*, where $aw = iw + sw$ is the *acceleration weight*. The settings of these coefficients greatly affect the behaviour of the swarm.

Equation (1) can also be formulated as in Equation (2):

$$\begin{cases} v_{ij}^{(t)} = w \cdot v_{ij}^{(t-1)} + \phi_i \cdot \left( pbest_{ij}^{(t-1)} - x_{ij}^{(t-1)} \right) + \phi_s \cdot \left( lbest_{ij}^{(t-1)} - x_{ij}^{(t-1)} \right) \\ 0 \le \left( \phi = \phi_i + \phi_s = iw \cdot U_{(0,1)} + sw \cdot U_{(0,1)} \right) \le \left( aw = iw + sw \right) \\ x_{ij}^{(t)} = x_{ij}^{(t-1)} + v_{ij}^{(t)} \end{cases} \quad (2)$$

where:

$\phi_i = iw \cdot U_{(0,1)} = U_{(0,iw)}$ : *Individual acceleration coefficient.*

$\phi_s = sw \cdot U_{(0,1)} = U_{(0,sw)}$ : *Social acceleration coefficient.*

$\phi = \left( \phi_i + \phi_s \right) \in \left[ 0, aw \right]$ : *Acceleration coefficient.*

Note that the *acceleration coefficients* in Equation (2) are the actual scaling factors of the difference vectors in the velocity update equation, whereas the *individuality* and *sociality weights* in Equation (1) are the upper limits of the scaling factors.

## 2.2 Neighbourhood

The cooperative, social behaviour that leads to the emergent optimization ability is given by the social network within the swarm; that is to say, by the definition of the neighbours which inform every particle of their best previous experiences. Thus the particle can be influenced by the best experience in its neighbourhood (**lbest**).

The social structure is typically referred to as *neighbourhood structure* or *neighbourhood topology*. Three classical neighbourhood topologies are shown in Figure 1, where the nodes in the graphs represent the particles, and the bidirectional edges stand for their interconnections.

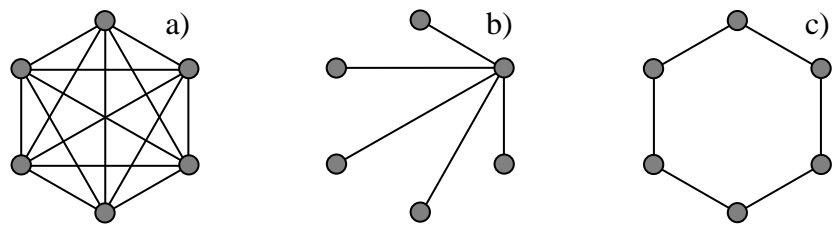

Figure 1: a) fully connected topology, where neighbourhood-size equals swarm-size; b) wheel topology, where neighbourhood-size equals swarm-size for one particle and two for the rest; c) ring topology with neighbourhood-size equal to three.

Numerous neighbourhood topologies can be found in the literature, and infinite ones can be thought of. Refer, for instance, to [16], [13], [17], and [18].

## 3 Coefficients

The settings of the coefficients in Equation (1) define the trajectories of the particles. Those of the original PSO algorithm, namely $w = 1$ and $iw = sw = 2$ ($aw = 4$), results in the particles performing a so-called *explosion*. That is to say, they diverge from the attractors. While limiting the size of the components of the particles' velocity does prevent the *explosion*, it does not help ensure or control the speed and form of convergence. Conversely, the settings of the coefficients in Equation (1) do.

## 3.1 Isolated particle

Ozcan and Mohan [8] [9] developed the first theoretical studies on the trajectory of a single, isolated, deterministic particle flying (surfing) over a one-dimensional space, pulled by stationary attractors. Clerc and Kennedy [11] went further and completely defined the system in a 5-dimenensional complex space, while also proposing constriction factors that ensure convergence. They generalized the system for continuous time, where the particle moves in real space for the integer parts of it only. Van den Bergh [10], Trelea [12] and Innocente [13] also carried out studies on the trajectories of the particle in this simplified system, obtaining −by different means− the same convergence region in the plane '$w$–$\phi$'. The studies offered in [13] lead to the convergence region shown in Figure 2. Thus only the pairs '$w$–$\phi$' within the shaded triangle ensure convergence (of the deterministic, isolated particle).

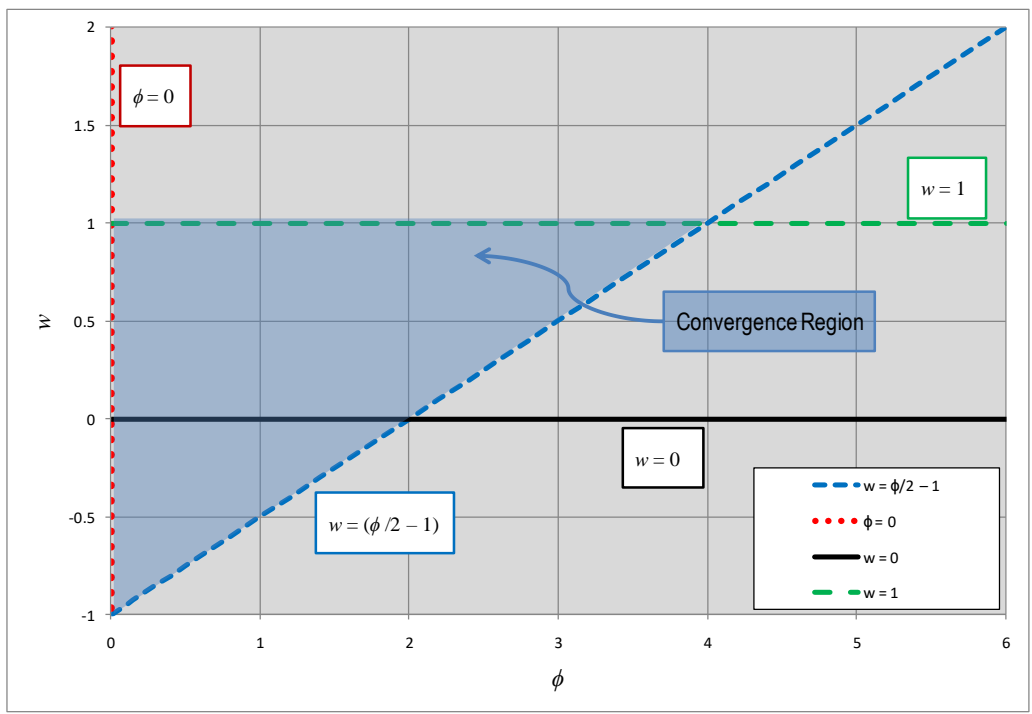

Figure 2: Convergence region in plane '$w$–$\phi$', with redundant bounds removed.

Figure 3 shows two examples of the trajectories resulting from pairs '$w$–$\phi$' within the convergence triangle for $w < 0$ and both attractors fixed at $x = 0$. Although convergence clearly takes place, values of $w < 0$ are −in general− disregarded, as the concept of carrying some inertia from the previous displacement would be disrupted. Thus the convergence region of practical interest is the gray area in Figure 4, where the region within the dotted parabola comprises the area where the roots of the characteristic polynomial of the recurrence relation of the deterministic particle's trajectory are complex conjugates (refer to [13]).

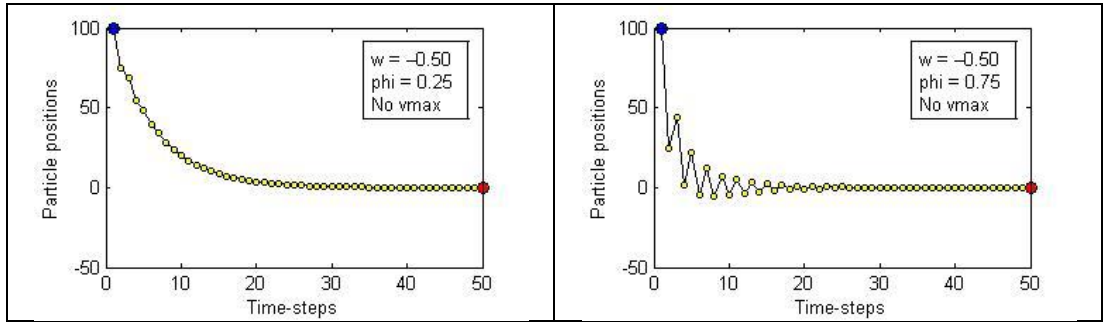

Figure 3: Two convergent trajectories of the deterministic particle for $w < 0$. On the left, $w = -0.50$ and $\phi = 0.25$ whereas on the right, $w = -0.50$ and $\phi = 0.75$.

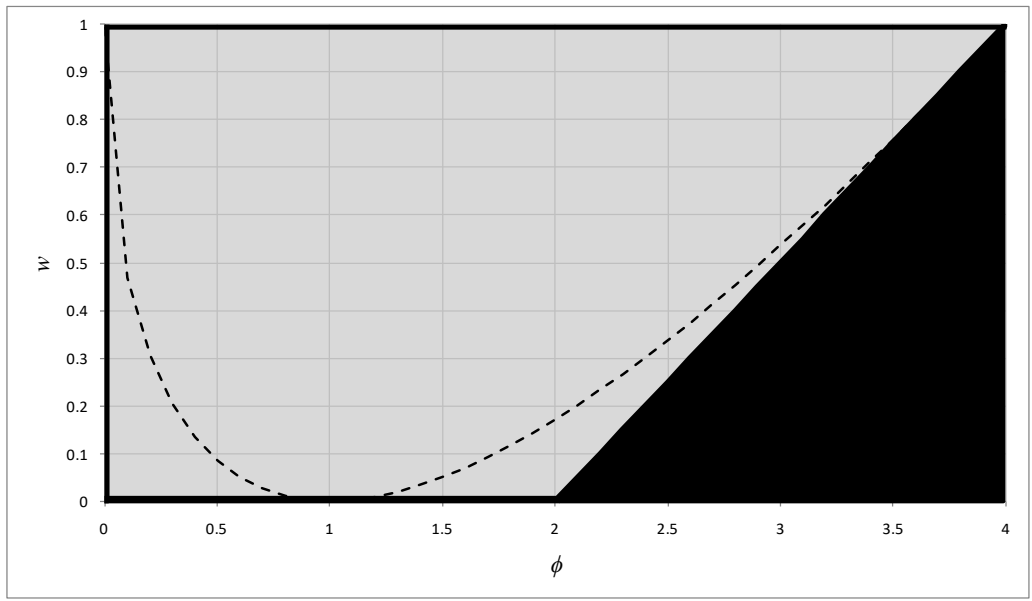

Figure 4: Area of practical interest of the convergence triangle in plane '$w$–$\phi$' (gray shade). The black triangle comprises points of the divergence region.

Ensured convergence is not the only important aspect of the particle's trajectory. The speed and form of convergence define the manner in which the search is carried out, and therefore have a critical impact on the final performance of the optimizer. Different combinations of $w$ and $\phi$ within the convergence region result in different amplitudes, frequencies, and speed of damping of the oscillations in the particle's trajectories.

20 pairs '$w$–$\phi$' resulting from all combinations of five values of $w$ and four values of $\phi$ are considered for a visual analysis of the particle's trajectory. These selected values are shown in Table 1 and Figure 5.

| | | $\phi$ | | | |
|---|---|---|---|---|---|
| | | **0.5** | **2** | **3.5** | **4** |
| | **1** | A1 | A2 | A3 | A4 |
| | **0.8** | B1 | B2 | B3 | B4 |
| $w$ | **0.5** | C1 | C2 | C3 | C4 |
| | **0.2** | D1 | D2 | D3 | D4 |
| | **0** | E1 | E2 | E3 | E4 |

Table 1: 20 selected pairs '$w$–$\phi$' to be considered for a visual analysis of the particle's trajectories. Note that only some of them are within convergence region.

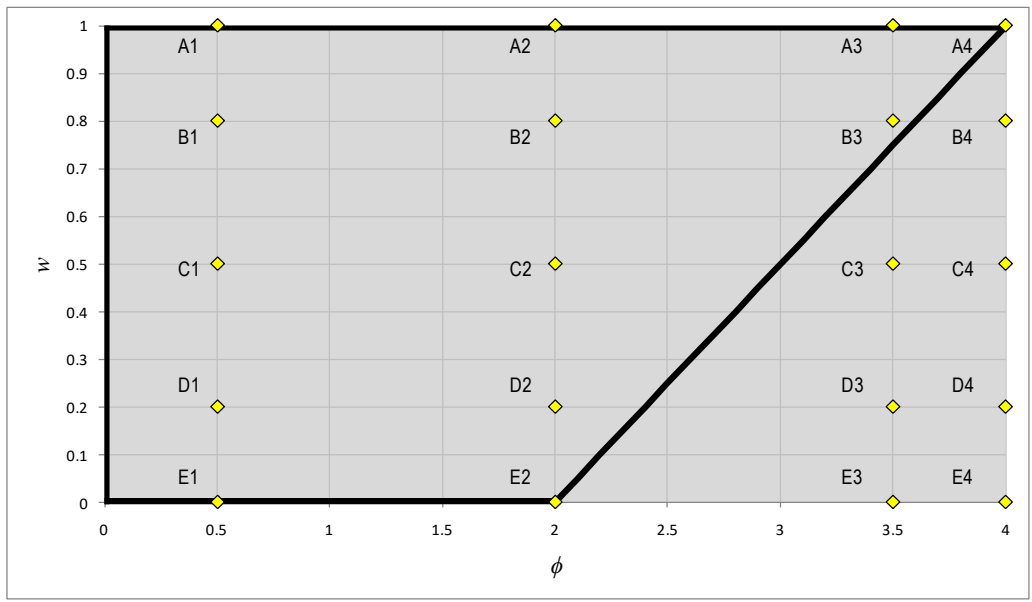

Figure 5: 20 selected pairs '$w$–$\phi$' to be considered for a visual analysis of the particle's trajectories. Note that only 8 of these pairs are within convergence region. The other 12 pairs lead to either (pseudo) cyclic or divergent trajectories.

The trajectories of the particle resulting from the 20 selected pairs '$w$–$\phi$' shown in Table 1 and Figure 5 are offered in Figure 6 and Figure 7, where both attractors are fixed at $x = 0$. Notice that Figure 6 shows the trajectories corresponding to the first two columns of points in Figure 5 whereas Figure 7 shows the trajectories corresponding to the other two columns. Hence, if these figures are assembled in the form [Figure 6 Figure 7], a grid of trajectories is obtained corresponding to their homologous pairs '$w$–$\phi$' in Figure 5.

It can be observed from Figure 6 and Figure 7 that the points within the convergence region (B1, B2, B3, C1, C2, D1, D2, and E1) effectively lead to convergence towards the attractor. Conversely, the points within the divergence region (B4, C3, C4, D3, D4, E3, E4) lead to the particle diverging from the attractor, where the lower the *inertia weight* ($w$) for a given *acceleration coefficient* ($\phi$) −and the greater the $\phi$ for a given $w$− the greater the explosion.

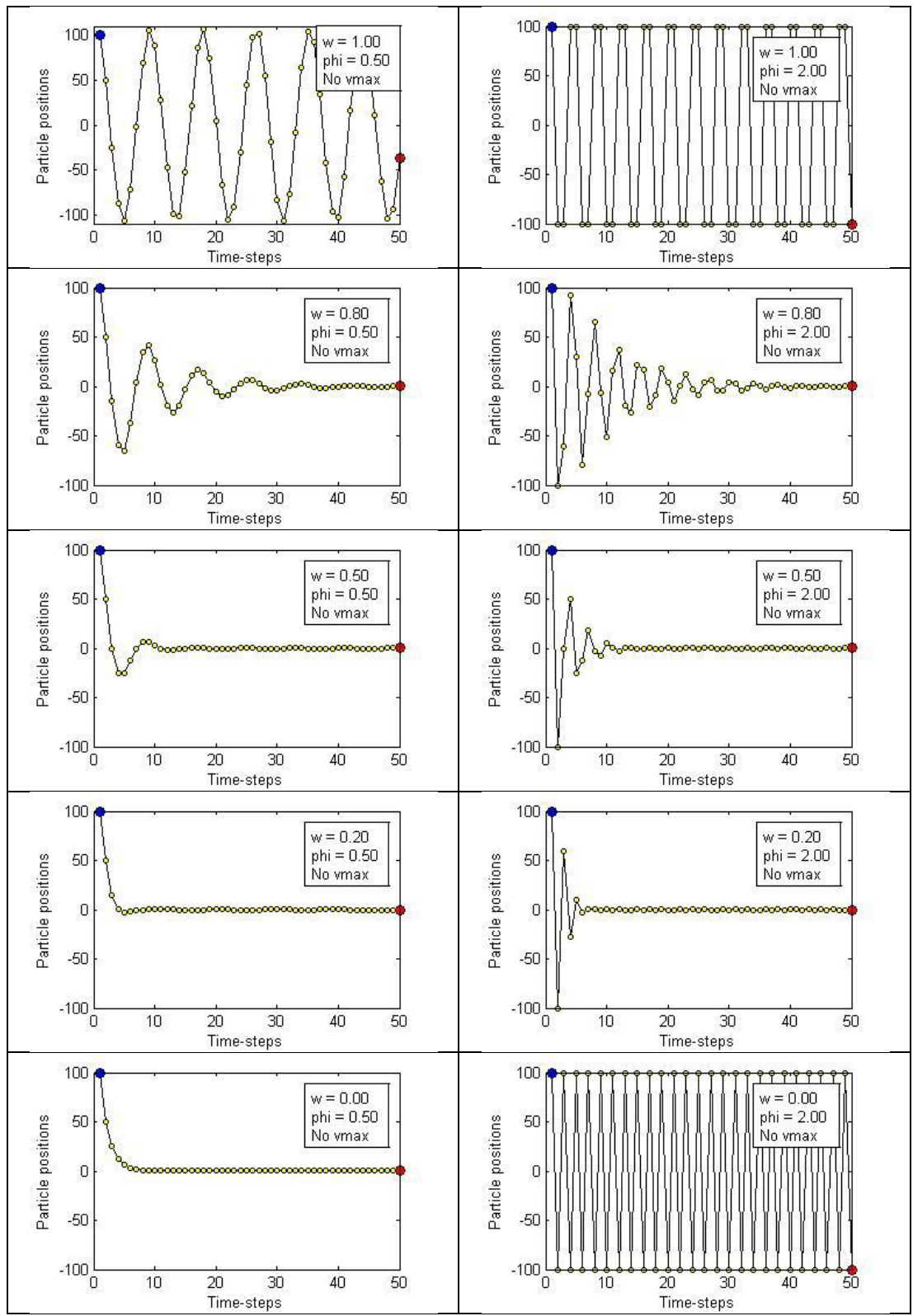

Figure 6: Trajectories of the deterministic particle with fixed attractors for the pairs A1 to E2 (first two columns) in Figure 5.

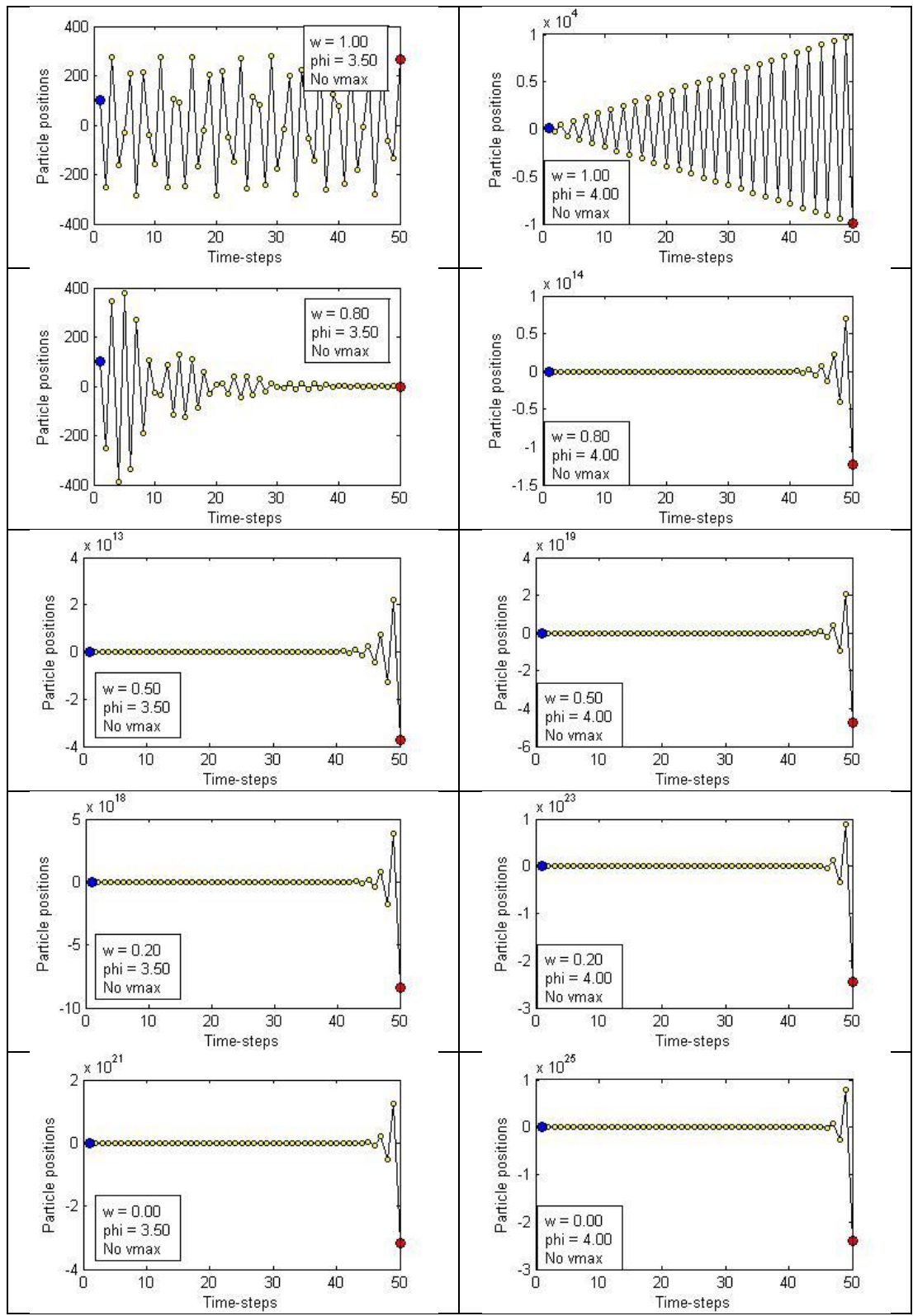

Figure 7: Trajectories of the deterministic particle with fixed attractors for the pairs A3 to E4 (last two columns) in Figure 5.

The points on the boundary –namely A1, A2, A3, A4, and E2– deserve a more extensive analysis. The important point is that they do not lead to convergence. According to the magnitudes of the roots of the characteristic polynomial associated to the recurrence relation of the particle's position, they may lead to cyclic, pseudo-cyclic, or divergent trajectories. Loosely speaking, points A1, A2 and A3 lead to (pseudo) cyclic trajectories because the roots are complex conjugates with a module equal to one; point A4 leads to a linearly divergent trajectory because both roots are real-valued and equal to '−1'; and point E2 leads to a cyclic trajectory because one root equals '0' and the other equals '−1' which results in the particle moving in opposite directions in consecutive time-steps while always keeping the same distance from the attractor. For a thorough analysis of this behaviour, refer to [13].

With regards to the different speeds of convergence, it can be observed that the closer to the boundaries of the convergence region the slower the convergence. Regarding the form of convergence, the greater the *inertia weight* ($w$) in relation to the *acceleration coefficient* ($\phi$) the lower the frequency of the oscillations in the trajectory. In other words, low $w$ with high $\phi$ leads to high frequencies –i.e. the attractor being overflown a higher number of times– whereas high $w$ with low $\phi$ results in low frequencies. Setting them both to small values is not advisable, as exploration is affected, while setting them both to great values −approaching point A4 in Figure 5− leads to slow convergence.

The analysis this far considered the particle as being pulled towards a single and stationary attractor, where $\phi$ is constant. The full PSO algorithm considers each particle pulled towards two distinct and dynamic attractors, where $\phi_i$ and $\phi_s$ (and therefore $\phi$) are random variables. Thus, in the full PSO algorithm, the particle is attracted towards a point that is a randomly weighed average of the individual and the social attractors in Equations (1) and (2). Hence the acceleration towards the individual attractor is weighed by a random number $\phi_i \in [0, iw]$ and the one towards the social attractor is weighed by a random number $\phi_s \in [0, sw]$. Therefore, it is not only the relative values of the *inertia weight* ($w$) and the *acceleration coefficient* ($\phi$) which affect the trajectories but also the relative values of $iw$ and $sw$, which award different importance to the individual and the social best previous experiences.

One would reasonably expect that higher $iw/sw$ ratios for a given *acceleration weight* ($aw = iw + sw$) would result in a more extensive exploration and a slower convergence, whereas lower ratios would quickly focus the search on small regions of the search-space thus performing a fine-grain search and increasing convergence speed. However, this does not necessarily happen, and even the opposite may occur.

To illustrate this, the trajectories corresponding to four consecutive runs of a particle initialized at $x = 100$ optimizing the 1-dimensional Sphere function with stationary social attractor and dynamic individual attractor initialized at $x = 90$, with $w = 0.70$ and $aw = 4.00$, and with randomness reintroduced into the algorithm, are presented in Figure 8 to Figure 10 for different $iw/sw$ ratios. Thus, the trajectories for $iw = aw$ (hence $sw = 0$) are shown in Figure 8; for $iw = 0.50 \cdot aw$ (hence $sw = iw$) in Figure 9; and for $iw = 0$ (hence $sw = aw$) in Figure 10.

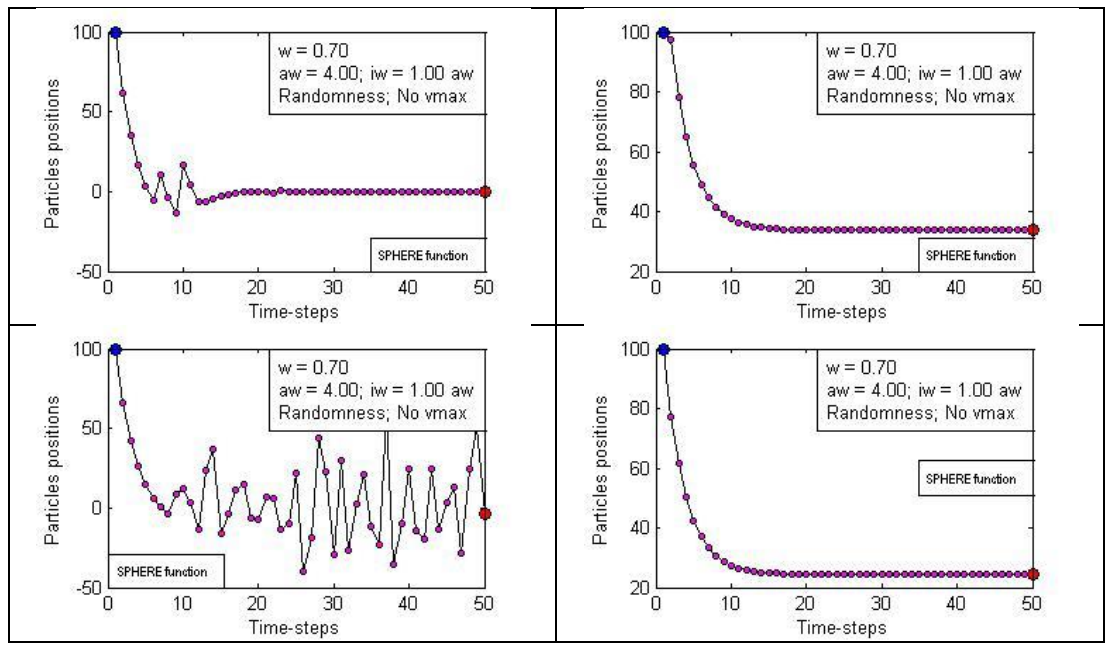

Figure 8: Trajectories corresponding to 4 consecutive runs of a particle initialized at $x = 100$, optimizing the 1-dimensional Sphere function, with stationary social attractor at $x = 0$ and dynamic individual attractor initialized at $x = 90$, for $w = 0.70$, $aw = 4.00$, and for $iw = aw$ (hence $sw = 0$) and randomness reintroduced.

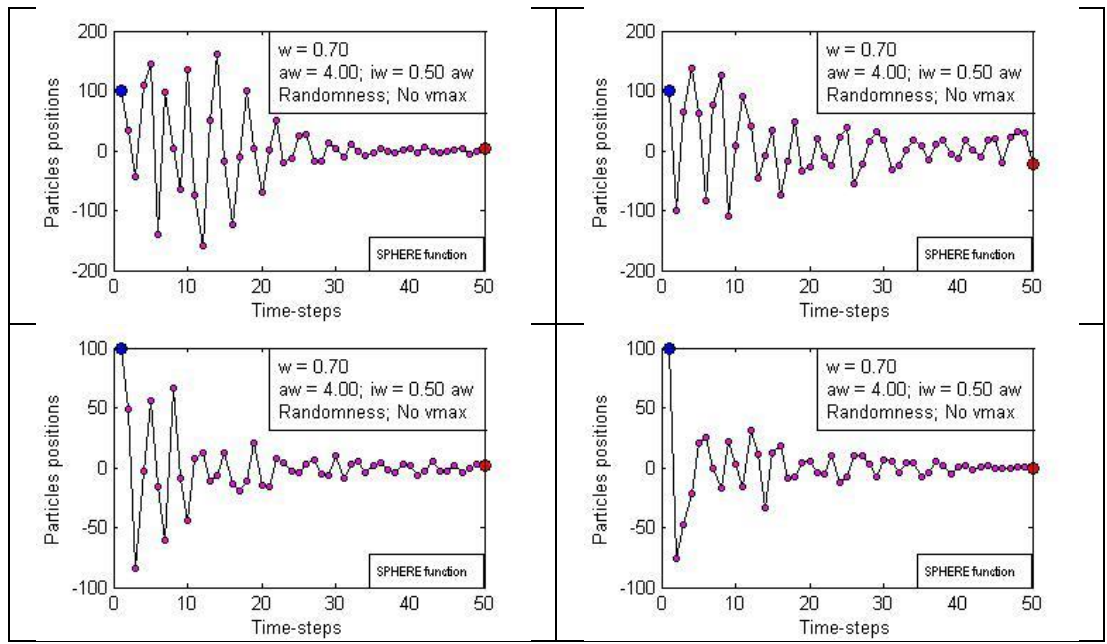

Figure 9: Trajectories corresponding to 4 consecutive runs of a particle initialized at $x = 100$, optimizing the 1-dimensional Sphere function, with stationary social attractor at $x = 0$ and dynamic individual attractor initialized at $x = 90$, for $w = 0.70$, $aw = 4.00$, and for $iw = 0.50 \cdot aw$ (hence $sw = iw$) and randomness reintroduced.

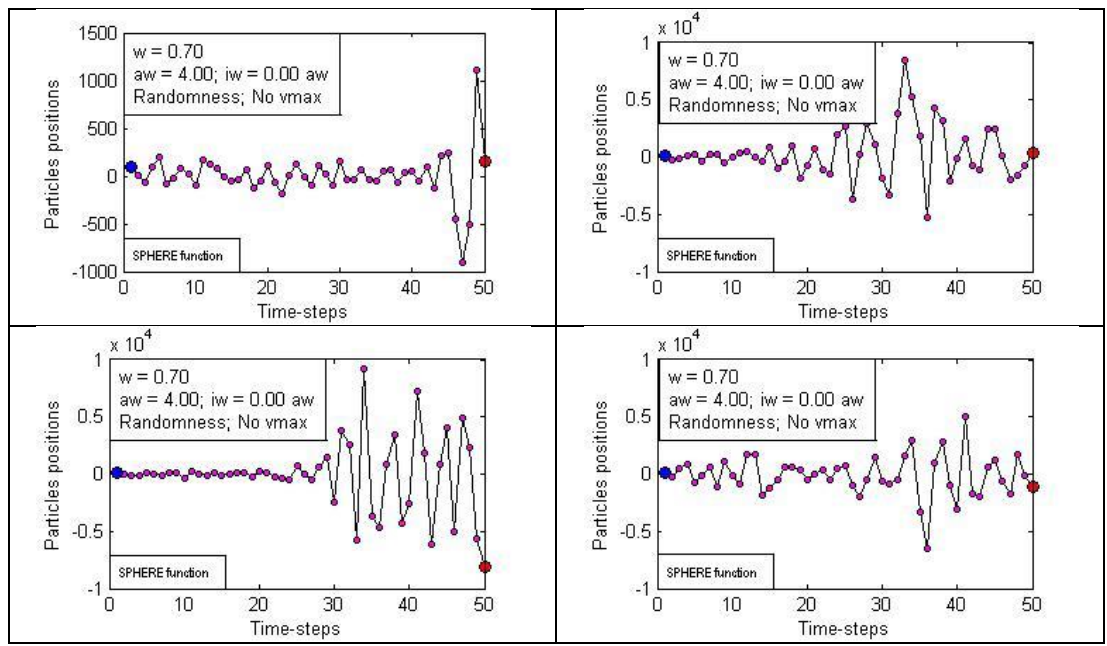

Figure 10: Trajectories corresponding to 4 consecutive runs of a particle initialized at $x = 100$, optimizing the 1-dimensional Sphere function, with stationary social attractor at $x = 0$ and dynamic individual attractor initialized at $x = 90$, for $w = 0.70$, $aw = 4.00$, and for $iw = 0$ (hence $sw = aw$) and randomness reintroduced.

As can be observed, setting very high $iw/sw$ ratios may lead the particle to gradually approach the social attractor in small step-sizes from one side only, rather than to a wide search of the space. In turn, setting very low $iw/sw$ ratios may even result in divergence rather than in enhancing the fine-grain search, as the step-sizes increase and the particle oscillates. Therefore, setting $iw$ and $sw$ to similar values, thus leaving the random weights in charge of dynamically altering their relative importance, results in a more desirable, reliable, and consistent behaviour.

## 3.2 Interacting particles

The trajectories shown in Figure 6 and Figure 7 correspond to a deterministic and isolated particle pulled by stationary attractors. The 20 '$w$–$\phi$' pairs analyzed were constant, while in the full PSO algorithm $\phi_i$, $\phi_s$ and $\phi$ vary randomly within the ranges shown in Equation (3).

$$\phi \in [0, aw]$$
$$\phi_i \in [0, iw]$$
$$\phi_s \in [0, sw]$$

(3)

The first step towards the full algorithm consists of reintroducing the random weights. Consider the horizontal line for $w = 0.80$ in Figure 5. Setting $aw = 3.50$ would mean that when $\phi = aw$ the behaviour is that of the point B3; when $\phi = 2.00$ the behaviour is that of point B2; and when $\phi = 0.50$ the behaviour is that of point B1. Therefore, analyzing the trajectories in Figure 6 and Figure 7 allows inferring that the swarm's behaviour would be rather explorative, and oscillatory with randomly varying frequencies. Similarly, consider the horizontal line for $w = 0.50$ in Figure 5, where the behaviour within the convergence region would be expected to be oscillatory, favouring a fine-grain search and faster convergence.

In order to show that these studies of an isolated particle also apply to a cooperative swarm, the trajectories of four interacting particles in a full PSO algorithm are shown for $w = 0.80$; $aw = 3.50$; and $iw = 0.50 \cdot aw$ in Figure 11; and for $w = 0.50$; $aw = 3.00$; and $iw = 0.50 \cdot aw$ in Figure 12. Clearly, exploration is more extensive in the former case, while convergence is notably faster in the latter.

Aiming to also observe whether the speculations on the $iw/sw$ ratios drawn for the isolated particle apply to a swarm of interacting particles, the trajectories for three additional ratios other than $iw/sw = 1.00$ are also offered. Thus Figure 13, Figure 14, and Figure 15 complement Figure 11, whilst Figure 16, Figure 17, and Figure 18 complement Figure 12. Comparing Figure 11, Figure 13, and Figure 14, it would appear that increasing $iw/sw$ ratios delay convergence. However, the trajectories are more erratic and less consistent for different runs than for the case of $iw/sw = 1.00$. In addition, the low $iw/sw$ ratio in Figure 15 ($0.20/0.80 = 0.25$) results in an even more erratic behaviour rather than in an increased convergence speed. Therefore, setting $iw = sw$ also appears convenient and safer in this case.

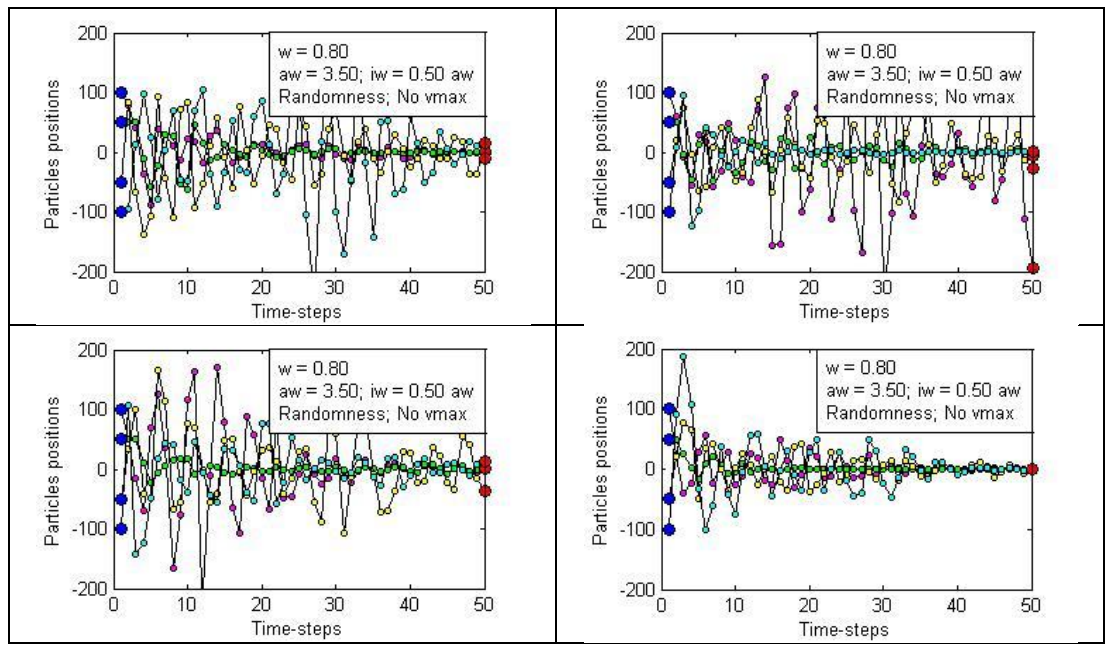

Figure 11: Trajectories corresponding to 4 consecutive runs of 4 particles initialized at $x = 100$, $x = 50$, $x = -50$, and $x = -100$, for $w = 0.80$, $aw = 3.50$ and $iw/sw = 1.00$, optimizing the 1-dimensional Sphere function. The initial individual best experiences coincide with the corresponding initial positions.

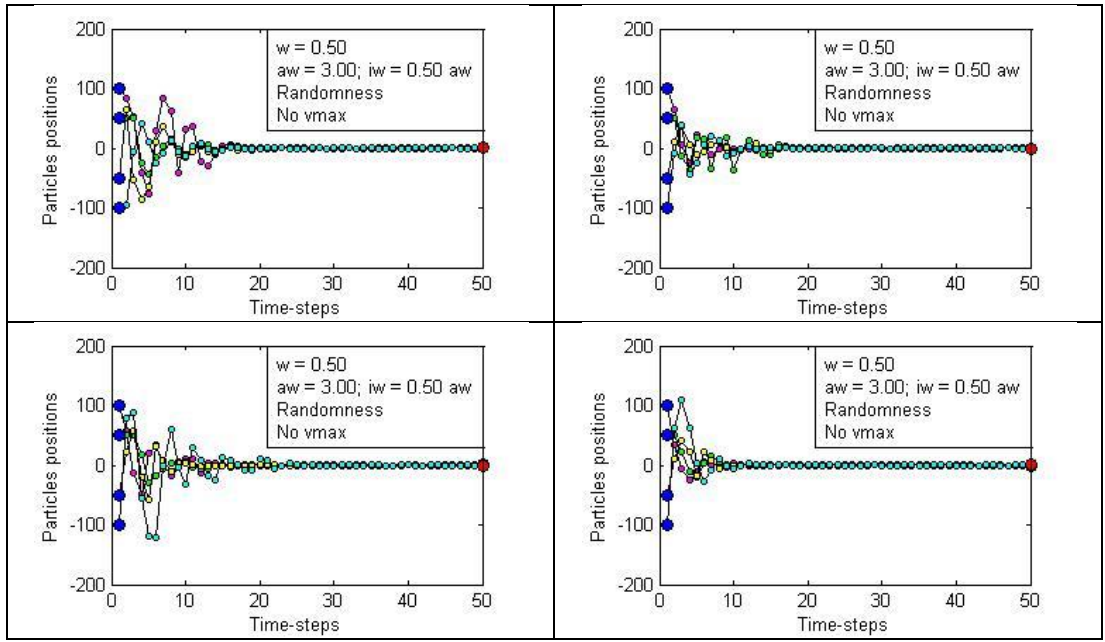

Figure 12: Trajectories corresponding to 4 consecutive runs of 4 particles initialized at $x = 100$, $x = 50$, $x = −50$, and $x = −100$, for $w = 0.50$, $aw = 3.00$ and $iw/sw = 1.00$, optimizing the 1-dimensional Sphere function.

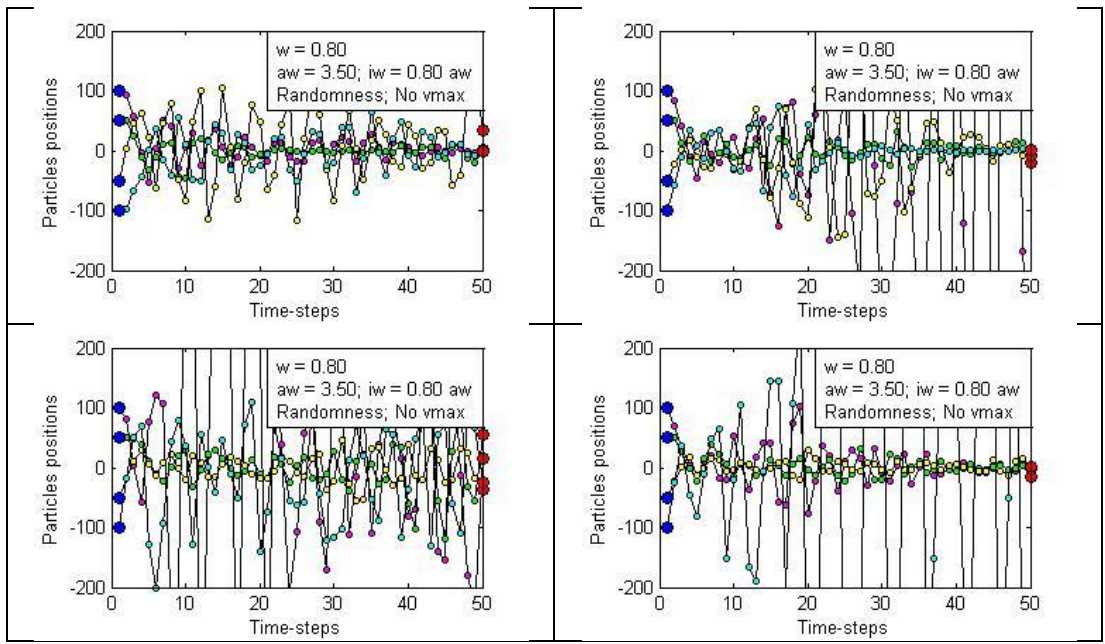

Figure 13: Trajectories corresponding to 4 consecutive runs of 4 particles initialized at $x = 100$, $x = 50$, $x = −50$, and $x = −100$, for $w = 0.80$, $aw = 3.50$ and $iw/sw = 4.00$, optimizing the 1-dimensional Sphere function.

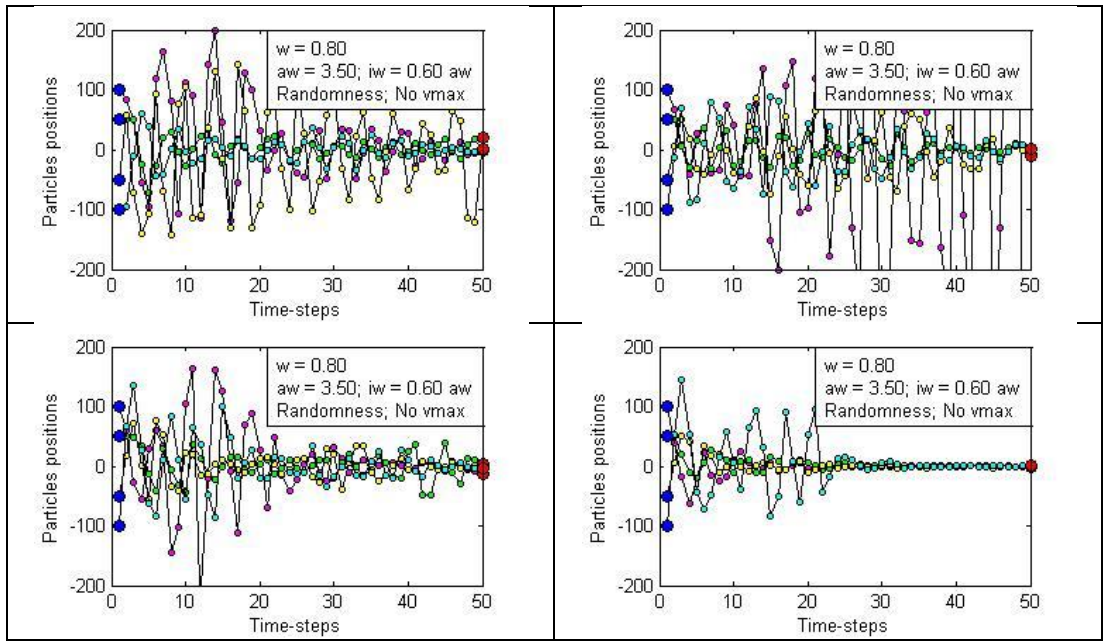

Figure 14: Trajectories corresponding to 4 consecutive runs of 4 particles initialized at $x = 100$, $x = 50$, $x = -50$, and $x = -100$, for $w = 0.80$, $aw = 3.50$ and $iw/sw = 1.50$, optimizing the 1-dimensional Sphere function.

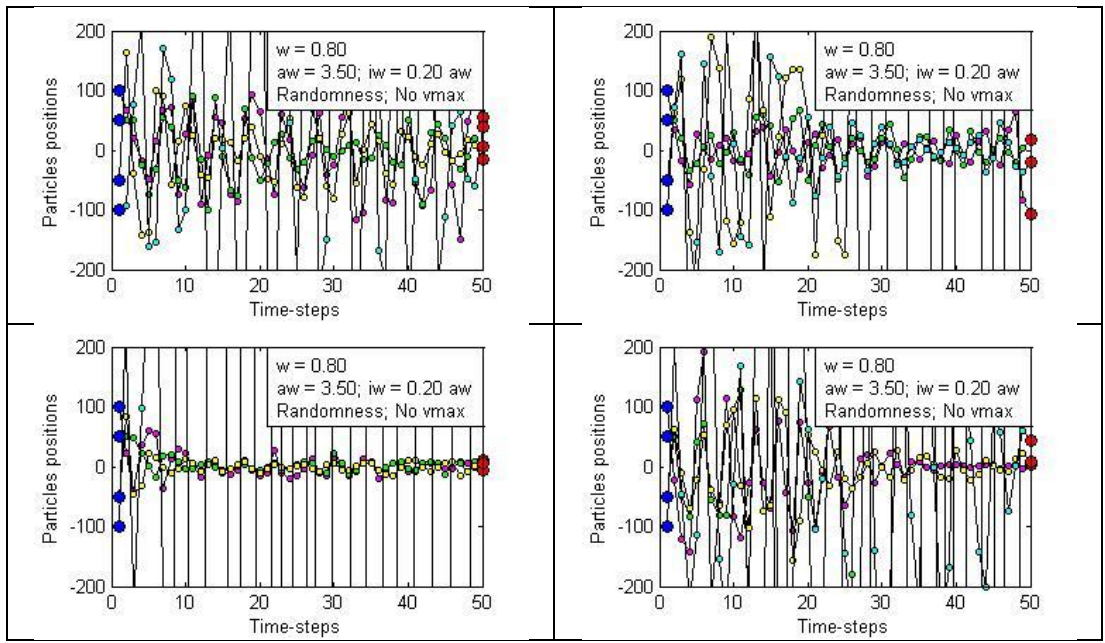

Figure 15: Trajectories corresponding to 4 consecutive runs of 4 particles initialized at $x = 100$, $x = 50$, $x = -50$, and $x = -100$, for $w = 0.80$, $aw = 3.50$ and $iw/sw = 0.25$, optimizing the 1-dimensional Sphere function.

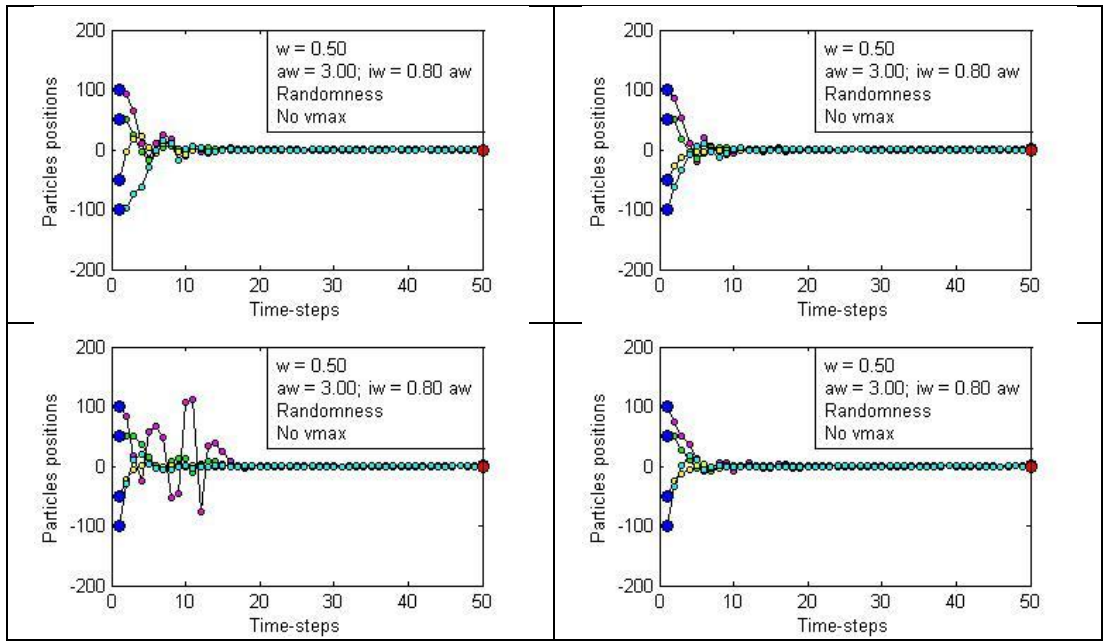

Figure 16: Trajectories corresponding to 4 consecutive runs of 4 particles initialized at $x = 100$, $x = 50$, $x = -50$, and $x = -100$, for $w = 0.50$, $aw = 3.00$ and $iw/sw = 4.00$, optimizing the 1-dimensional Sphere function.

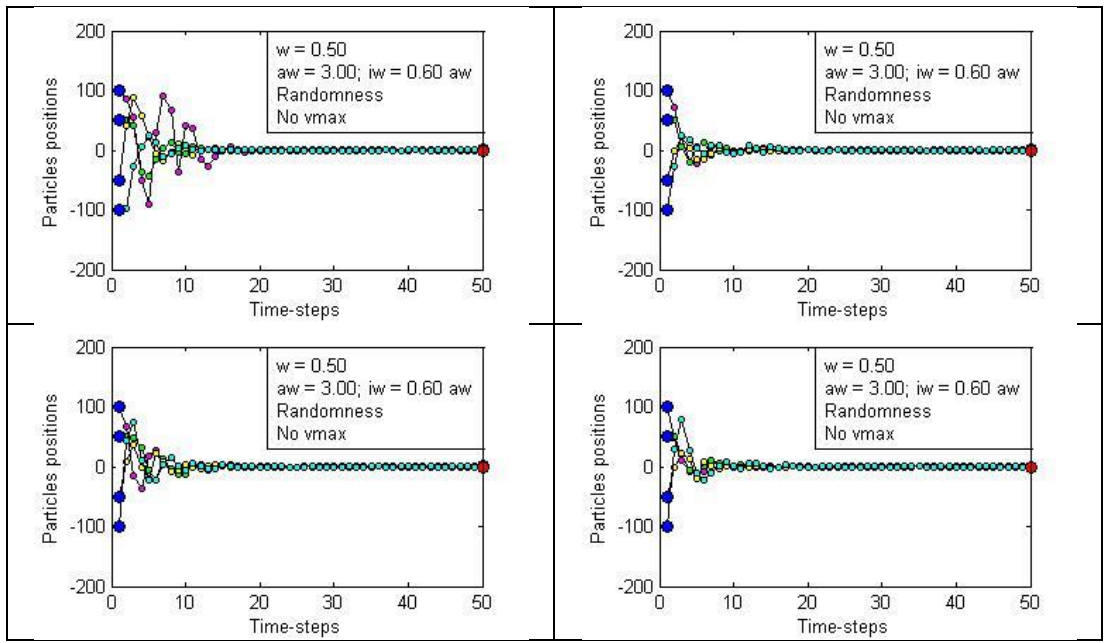

Figure 17: Trajectories corresponding to 4 consecutive runs of 4 particles initialized at $x = 100$, $x = 50$, $x = -50$, and $x = -100$, for $w = 0.50$, $aw = 3.00$ and $iw/sw = 1.50$, optimizing the 1-dimensional Sphere function.

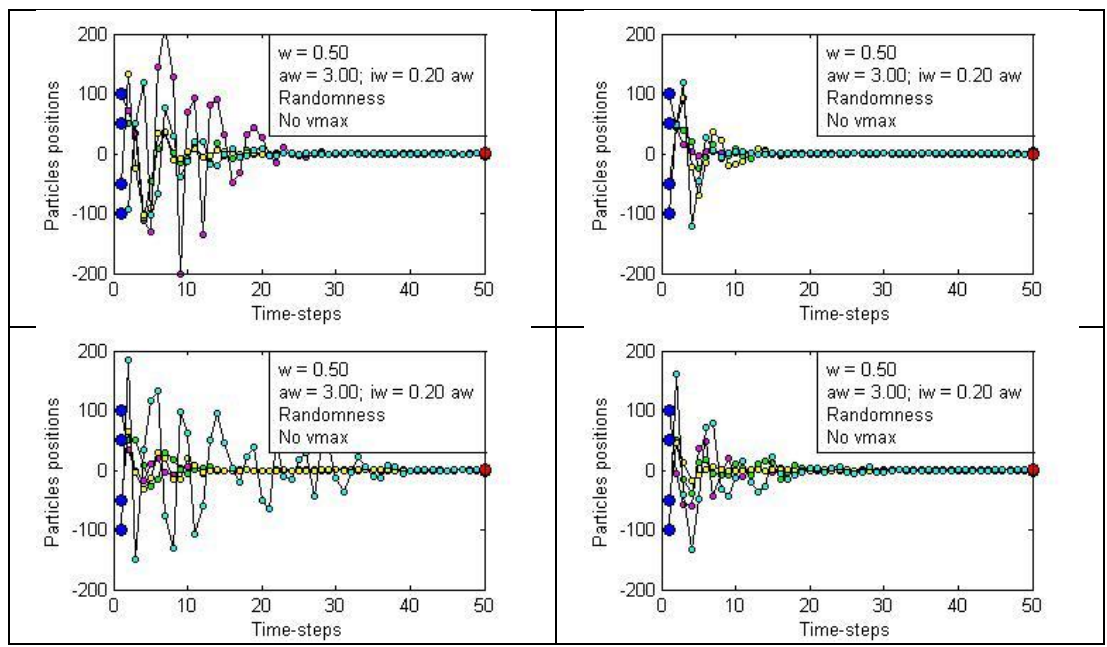

Figure 18: Trajectories corresponding to 4 consecutive runs of 4 particles initialized at $x = 100$, $x = 50$, $x = -50$, and $x = -100$, for $w = 0.50$, $aw = 3.00$ and $iw/sw = 0.25$, optimizing the 1-dimensional Sphere function.

Observing Figure 12, Figure 16, Figure 17, and Figure 18, it shows that the speed of convergence increases with increasing $iw/sw$ ratios while decreasing ratios result in slightly higher exploration. Again, $iw/sw = 1$ appears to be the safest choice.

Figure 12 shows that setting $aw$ within the convergence region in Figure 5 for an inertia weight as low as 0.50 already leads to the search quickly focusing on a small region of the space. Even smaller values are not advisable as the space would be practically unexplored. Aiming to illustrate this, the trajectories corresponding to four consecutive runs of four particles with $w = 0.20$, $aw = 2.40$ and $iw/sw = 1.00$ are offered in Figure 19.

## 4    Neighbourhood Topology

Since the aim of this paper is not to present a thorough study of neighbourhood topologies in PSO but to illustrate the combined effect of the coefficients' settings and the neighbourhood structure in the convergent social behaviour of the particles, only three classical topologies are considered: the *global* topology, the *wheel* topology, and the *ring* topology with 2 neighbours. The first one favours fast convergence whereas the last one favours reluctance to getting stuck in poor sub-optimal solutions. The wheel topology is intermediate. For further, more extensive studies on neighbourhood topologies, refer to [16], [13], [17], and [18].

The objective here is to address the question of how much of the speed and form of convergence should be controlled by the settings of the coefficients and how much by the structure of the neighbourhoods. For instance, an explorative behaviour may be obtained by setting coefficients that discourage too fast a convergence, or by setting coefficients that favour fast convergence but delaying the speed of spread of information throughout the swarm by reducing the number of interconnections between the particles in the neighbourhood. The former case comprises a more traditional form of exploration while in the latter case exploration is carried out by exploitation of numerous local best experiences.

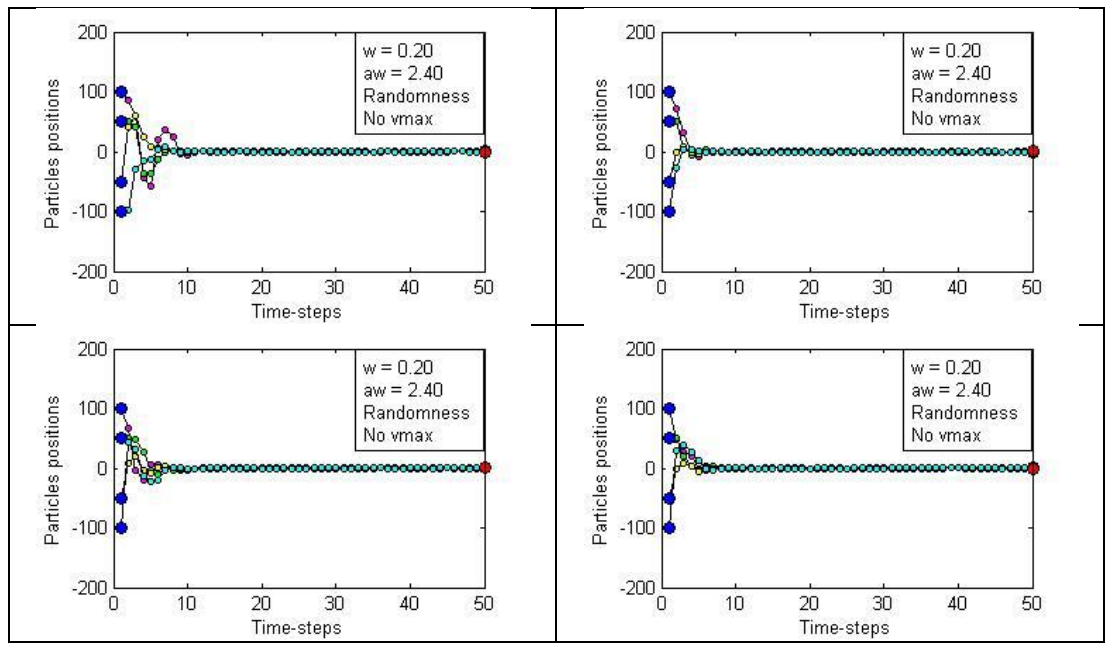

Figure 19: Trajectories corresponding to 4 consecutive runs of 4 particles initialized at $x = 100$, $x = 50$, $x = -50$, and $x = -100$, for $w = 0.20$, $aw = 2.40$ and $iw/sw = 1.00$, optimizing the 1-dimensional Sphere function.

Thus, 12 experiments are performed: one for each combination of a set of settings favouring a more extensive exploration ($w = 0.80$; $aw = 3.50$ and $iw/sw = 1.00$); a set of settings favouring fine-grain search ($w = 0.50$; $aw = 3.00$ and $iw/sw = 1.00$); the three aforementioned neighbourhood topologies; and two benchmark functions. Only two problems, namely the Sphere function (unimodal) and the Rastrigin function (multimodal), are considered because the aim is not to carry out an extensive numerical study but to illustrate how these two important aspects of PSO −the settings of the coefficients and the neighbourhood topology− work together to control the speed and form of convergence.

# 5    Experimental Results

The details of the implementation for the experiments are as follows: 50 particles; 10000 time-steps; synchronous update of the best experiences; the best of 1000 Latin Hypercube samplings for the particles' initialization (according to the maximum minimum distance criterion); velocities initialized to zero; and the best individual experiences initialized so that every component of each of them is at the same distance from the corresponding component of the homologous initial position. Six PSO algorithms are used in the experiments, which differ from each other only in the features described in the following:

PSO1:  $w = 0.80$;    $iw = sw = 1.75$ ($aw = 3.50$);  *global* topology
PSO2:  $w = 0.80$;    $iw = sw = 1.75$ ($aw = 3.50$);  *wheel* topology
PSO3:  $w = 0.80$;    $iw = sw = 1.75$ ($aw = 3.50$);  *ring* topology with 2 neighbours
PSO4:  $w = 0.50$;    $iw = sw = 1.50$ ($aw = 3.00$);  *global* topology
PSO5:  $w = 0.50$;    $iw = sw = 1.50$ ($aw = 3.00$);  *wheel* topology
PSO6:  $w = 0.50$;    $iw = sw = 1.50$ ($aw = 3.00$);  *ring* topology with 2 neighbours

The main results obtained by these 6 particle swarm optimizers are offered in Table 2, where 25 runs are performed for the statistics. The intermediate results by the time 10% of the total length of the search is carried out (1000 time-steps) are also presented in the table, while the convergence curves of the mean best solution found are offered in Figure 20 to Figure 23.

The first obvious observation from the convergence curves is that the settings of the coefficients in PSO1 to PSO3 lead to remarkably slower convergence than those in PSO4 to PSO6 for both problems tested. In fact, convergence is not even achieved after 10000 time-steps by the PSO4 to PSO6 for the multimodal 30-dimensional Rastrigin function (see Figure 22). Note that this is in agreement with the studies carried out in section 3.

The complete stagnation of the curves in Figure 21 −where the conflict function is unimodal− also illustrates how PSO may converge to any point, which does not need to be optimal in any sense.

Comparing the different neighbourhood structures, the global topology is the one leading to the fastest convergence in Figure 20 and Figure 23, as it is to be expected. While no topology reached convergence in Figure 22, the global topology is still the one reaching smaller values of the conflict function. Should the search be extended, it would be expectable that improvement stagnates sooner for the global topology. A bit of a strange case can be observed in Figure 21, where the ring topology with two neighbours converges faster than the global topology. This is because the problem is unimodal and improvement is more or less constant. Hence the locality of the ring topology does not sensibly delay clustering, as the individual experiences of the particles and the local social attractors successively move closer to the global optimum. Since the particles for the ring topology tend to move in smaller steps whereas the ones for the global topology have greater momentum −given the greater accelerations due to the greater distances to the only social attractor−, convergence ends up being faster for the local ring topology. Nevertheless, convergence is still very fast for the global topology in Figure 21 as well.

Surprisingly, the wheel topology consistently exhibits slower convergence of the mean best solution than the ring topology with 2 neighbours. This is probably due to the fact that the spread of acquired good information heavily depends on the performance of the centre of the wheel (one fixed particle). In fact, note that the best and the median solutions found by the wheel topology are not consistently the worst among all three topologies as the mean and worst solutions are (refer to Table 2).

| PROBLEM | Optimum | Coefficients | Neighbourhood | Time-steps | BEST | MEDIAN | MEAN | WORST | Mean FEs |
|---|---|---|---|---|---|---|---|---|---|
| 30D SPHERE | 0.000000 | $w = 0.80$ $aw = 3.50$ | GLOBAL | 1.00E+03 | 84.270484 | 160.092706 | 186.843878 | 437.604121 | 5.00E+04 |
| | | | | **1.00E+04** | **0.000000** | **0.000014** | **0.000084** | **0.000951** | **5.00E+05** |
| | | | WHEEL | 1.00E+03 | 247.525718 | 1081.719133 | 1077.135417 | 2700.436182 | 5.00E+04 |
| | | | | **1.00E+04** | **0.021866** | **2.233853** | **4.849282** | **28.834863** | **5.00E+05** |
| | | | RING (2 neighbours) | 1.00E+03 | 220.875113 | 658.754415 | 633.619375 | 1160.972962 | 5.00E+04 |
| | | | | **1.00E+04** | **0.004287** | **0.037783** | **0.072970** | **0.243672** | **5.00E+05** |
| | | $w = 0.50$ $aw = 3.00$ | GLOBAL | 1.00E+03 | 20.390171 | 189.727825 | 235.425463 | 913.183162 | 5.00E+04 |
| | | | | **1.00E+04** | **20.390171** | **189.727825** | **235.425463** | **913.183162** | **5.00E+05** |
| | | | WHEEL | 1.00E+03 | 19.001953 | 179.217902 | 382.267190 | 1678.237363 | 5.00E+04 |
| | | | | **1.00E+04** | **2.464484** | **158.996078** | **260.997827** | **1056.028231** | **5.00E+05** |
| | | | RING (2 neighbours) | 1.00E+03 | 0.000000 | 0.000000 | 0.000000 | 0.000000 | 5.00E+04 |
| | | | | **1.00E+04** | **0.000000** | **0.000000** | **0.000000** | **0.000000** | **5.00E+05** |
| 30D RASTRIGIN | 0.000000 | $w = 0.80$ $aw = 3.50$ | GLOBAL | 1.00E+03 | 51.586707 | 120.077491 | 119.467904 | 174.911883 | 5.00E+04 |
| | | | | **1.00E+04** | **13.954606** | **27.176228** | **27.486260** | **45.510433** | **5.00E+05** |
| | | | WHEEL | 1.00E+03 | 128.656522 | 174.521037 | 173.897292 | 233.670283 | 5.00E+04 |
| | | | | **1.00E+04** | **26.563912** | **49.047826** | **58.012713** | **164.470723** | **5.00E+05** |
| | | | RING (2 neighbours) | 1.00E+03 | 92.465587 | 123.533302 | 127.661820 | 164.759923 | 5.00E+04 |
| | | | | **1.00E+04** | **22.700181** | **42.126203** | **41.153665** | **57.174744** | **5.00E+05** |
| | | $w = 0.50$ $aw = 3.00$ | GLOBAL | 1.00E+03 | 49.264311 | 71.737281 | 75.252837 | 119.428284 | 5.00E+04 |
| | | | | **1.00E+04** | **49.264311** | **71.737281** | **75.251483** | **119.428284** | **5.00E+05** |
| | | | WHEEL | 1.00E+03 | 49.198498 | 75.887040 | 76.339189 | 111.487226 | 5.00E+04 |
| | | | | **1.00E+04** | **48.826491** | **72.757300** | **74.812050** | **111.440417** | **5.00E+05** |
| | | | RING (2 neighbours) | 1.00E+03 | 36.650971 | 56.972384 | 60.162610 | 86.968062 | 5.00E+04 |
| | | | | **1.00E+04** | **35.818501** | **55.717617** | **56.831956** | **85.566267** | **5.00E+05** |

Table 2: Statistical results obtained for the 30-dimensional Sphere (unimodal) and the 30-dimensional Rastrigin (multimodal) functions by a PSO algorithm with three different neighbourhood topologies and two different coefficients' settings. 25 runs are performed for the statistics.

Finally, the most important analysis is that of the combined effect of coefficients' settings and neighbourhood topologies in controlling the speed and form of convergence. Thus, it can be observed that if a rather explorative coefficients' setting is selected (PSO1 to PSO3 in Figure 20 and Figure 22), a neighbourhood topology that favours rather fast spread of information throughout the swarm –the global topology being the fastest− is advisable for a balanced trade-off between exploration and exploitation. Conversely, if a rather exploitative coefficients' setting is selected (PSO4 to PSO6 in Figure 21 and Figure 23), a more local neighbourhood topology is advisable in order to avoid premature convergence (e.g. ring topology with only two neighbours per particle). Note that an exploitative coefficients' setting coupled with the global topology in Figure 21 and Figure 23 led to premature convergence, whereas an explorative coefficients' setting coupled with a very local ring topology with 2 neighbours in Figure 20 and Figure 22 excessively delayed convergence. Of course it would be reasonable to expect that, if the search was extended, better solutions would be found eventually.

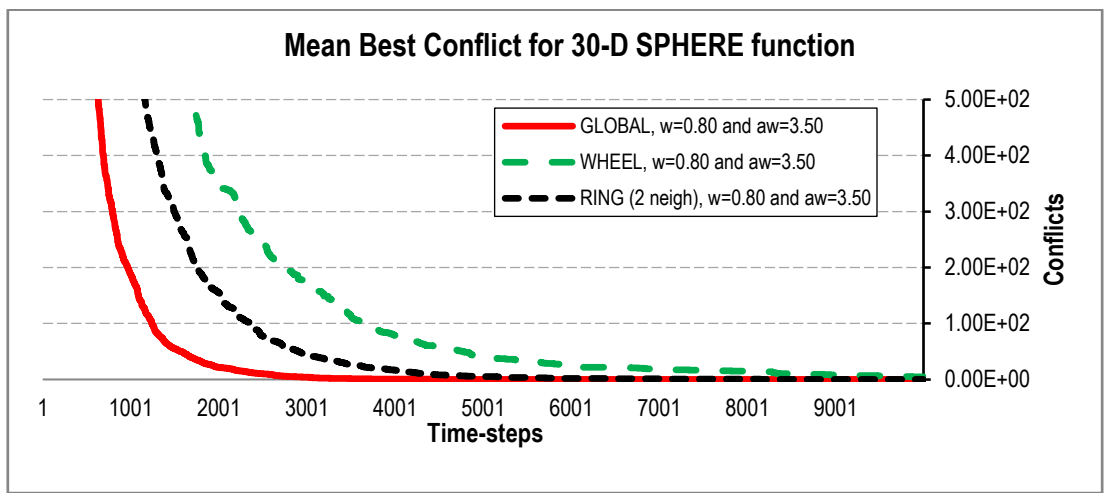

Figure 20: Convergence curves of the mean best conflict found by a PSO algorithm with $w = 0.80$, $aw = 3.50$ and $iw/sw = 1.00$ for the 30-dimensional Sphere function. Three different neighbourhood topologies are tested.

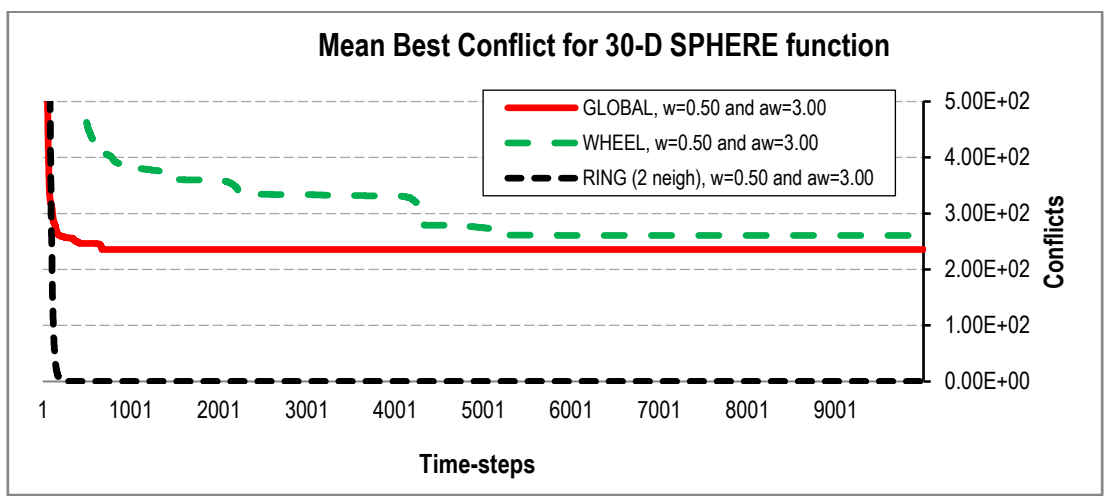

Figure 21: Convergence curves of the mean best conflict found by a PSO algorithm with $w = 0.50$, $aw = 3.00$ and $iw/sw = 1.00$ for the 30-dimensional Sphere function. Three different neighbourhood topologies are tested.

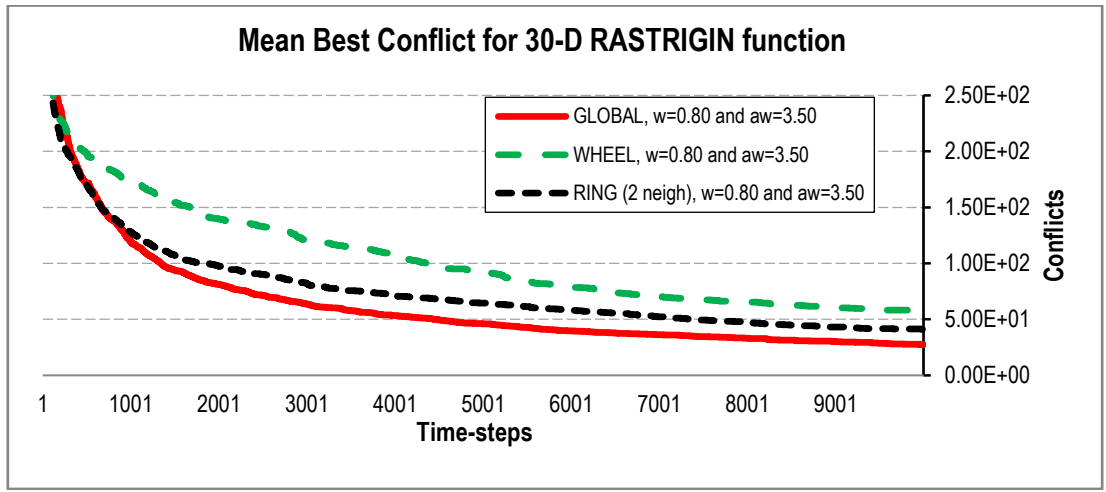

Figure 22: Convergence curves of the mean best conflict found by a PSO algorithm with $w = 0.80$, $aw = 3.50$ and $iw/sw = 1$ for the 30-dimensional Rastrigin function. Three different neighbourhood topologies are tested.

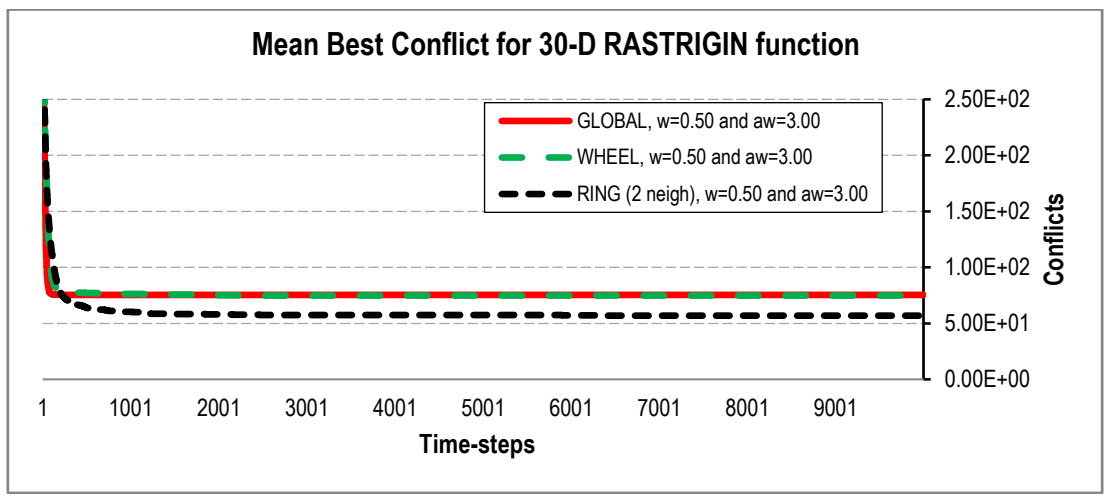

Figure 23: Convergence curves of the mean best conflict found by a PSO algorithm with $w = 0.50$, $aw = 3.00$ and $iw/sw = 1$ for the 30-dimensional Rastrigin function. Three different neighbourhood topologies are tested.

## 6    Conclusions

The aim of this paper was two-fold: first to offer some guidelines on the impact of different coefficients' settings in the speed and form of convergence; and second to illustrate the combined effect of the neighbourhood topology and the coefficients' settings on the performance of the PSO algorithm.

Thus, the convergence region of the '$w$–$\phi$' plane was presented, and the effect on the trajectory of a single deterministic and isolated particle was analyzed for different sub-regions in terms of the speed and form of convergence. The effect of setting the individuality and sociality weights to different values for a given acceleration weight was also explored. Experiments were carried out for a small swarm of four particles and the 1-dimensional Sphere function in order to analyze the trajectories and to observe whether the conclusions derived from the study of the isolated, deterministic particle still hold for the full PSO algorithm. It appears that, in a general sense, they do.

Finally, a set of experiments on two 30-dimensional problems –namely the unimodal Sphere function and the multimodal Rastrigin function− were performed for all the combinations of two sets of coefficients' settings and three neighbourhood topologies: a set of coefficients' settings favouring exploration; a set of coefficients' settings favouring fine-grain search; the global topology; the wheel topology; and the ring topology with 2 neighbours per particle. The results and convergence curves of these experiments showed the effect that the coefficients, the neighbourhood topologies, and their different combinations have on the performance of the optimizer. Thus the user can decide upon the coefficients and the neighbourhoods according to the type of search desired.